\documentclass[iop,numberedappendix]{emulateapj-rtx4}
\usepackage{graphicx,natbib,color,bm,url,times}
\graphicspath{{./fig/}{./png/}}

\newcommand{\EQ}{\begin{equation}}
\newcommand{\EN}{\end{equation}}
\newcommand{\EQA}{\begin{eqnarray}}
\newcommand{\ENA}{\end{eqnarray}}

\newcommand{\Eq}[1]{Equation~(\ref{#1})}
\newcommand{\Eqs}[2]{Equations~(\ref{#1}) and~(\ref{#2})}

\newcommand{\Sec}[1]{Section~\ref{#1}}

\newcommand{\Fig}[1]{Figure~\ref{#1}}

\newcommand{\Figp}[2]{Figure~\ref{#1}({#2})}

\newcommand{\bra}[1]{\langle #1\rangle}

{}
{}
{}

{}
{}
{}
{}
{}
{}
{}
{}
{}
{}
{}
{}
{}
{}
{}
{}
{}

{}

{}

{}
{}
{}

\newcommand{\hatB}{\hat{B}}

\newcommand{\meanAA}{{\overline{\bm{A}}}}
\newcommand{\meanBB}{{\overline{\bm{B}}}}

\newcommand{\kk}{\bm{k}}
\newcommand{\KK}{\bm{K}}

\newcommand{\xx}{\bm{x}}
\newcommand{\XX}{\bm{X}}
\newcommand{\aaaa}{\bm{a}}
\newcommand{\bb}{\bm{b}}

\newcommand{\BB}{\bm{B}}

\newcommand{\AAA}{\bm{A}}

\newcommand{\nab}{{\bm{\nabla}}}

\newcommand{\xxi}{\mbox{\boldmath $\xi$} {}}

\newcommand{\ii}{{\rm i}}

\newcommand{\dd}{{\rm d} {}}

\def\degr{\hbox{$^\circ$}}

\def\Pem{\mbox{\rm Pr}_{\rm M}}
\def\Rm{R_{\rm m}}

\def\Imag{\mbox{\rm Im}}

\def\half{{\textstyle{1\over2}}}

\newcommand{\G}{\,{\rm G}}

\newcommand{\cm}{\,{\rm cm}}

\newcommand{\Mm}{\,{\rm Mm}}

\newcommand{\erg}{\,{\rm erg}}

\newcommand{\etal}{et al.}

\newcommand{\yaj}[3]{ #1, {AJ,} {#2}, #3}

\newcommand{\yapj}[3]{ #1, {ApJ,} {#2}, #3}

\newcommand{\yapjl}[3]{ #1, {ApJ,} {#2}, #3}

\newcommand{\yan}[3]{ #1, {Astron.\ Nachr.,} {#2}, #3}

\newcommand{\yana}[3]{ #1, {A\&A,} {#2}, #3}

\newcommand{\yanar}[3]{ #1, {A\&A Rev.,} {#2}, #3}

\newcommand{\yjfm}[3]{ #1, {J.\ Fluid Mech.,} {#2}, #3}

\newcommand{\yjetp}[3]{ #1, {Sov.\ Phys.\ JETP,} {#2}, #3}

\newcommand{\yanf}[3]{ #1, {Ann. Rev. Fluid Mech.,} {#2}, #3}
\newcommand{\yrpp}[3]{ #1, {Rep.\ Prog.\ Phys.,} {#2}, #3}

\newcommand{\yprl}[3]{ #1, {Phys.\ Rev.\ Lett.,} {#2}, #3}

\newcommand{\ymn}[3]{ #1, {MNRAS,} {#2}, #3}
\newcommand{\ynat}[3]{ #1, {Nature,} {#2}, #3}

\newcommand{\ysph}[3]{ #1, {Solar Phys.,} {#2}, #3}

\newcommand{\yprd}[3]{ #1, {Phys.\ Rev.\ D,} {#2}, #3}
\newcommand{\ypre}[3]{ #1, {Phys.\ Rev.\ E,} {#2}, #3}

\newcommand{\yssr}[3]{ #1, {Spa.\ Sci.\ Rev.,} {#2}, #3}

\newcommand{\yjour}[4]{ #1, {#2}, {#3}, #4}

\newcommand{\ybook}[3]{ #1, {#2} (#3)}

\newcommand{\sapj}[2]{ #1, {ApJ}, submitted, arXiv:#2}

\newcommand{\ypisoe}[3]{#1, Proc. Int. Soc. Opt. Eng., {#2}, {#3}}
\begin{document}


\title{Bihelical spectrum of solar magnetic helicity and its evolution}
\author{Nishant K. Singh$^1$, Maarit J. K\"apyl\"a$^{1,2}$,
Axel Brandenburg$^{3,4,5,6}$, Petri J. K\"apyl\"a$^{7,2,1}$,
Andreas Lagg$^1$ and \\ Ilpo Virtanen$^8$
}
\affil{
$^1$Max Planck Institute for Solar System Research,
              Justus-von-Liebig-Weg 3, D-37077 G\"ottingen, Germany,\\
$^2$ReSoLVE Centre of Excellence, Department of Computer Science, 
	      Aalto University, PO Box 15400, FI-00076 Aalto, Finland\\
$^3$ NORDITA, KTH Royal Institute of Technology and Stockholm University, 
              Roslagstullsbacken 23, SE-10691 Stockholm, Sweden\\
$^4$ Department of Astronomy, AlbaNova University Center,
              Stockholm University, SE-10691 Stockholm, Sweden\\
$^5$ JILA and Department of Astrophysical and Planetary Sciences,
              Box 440, University of Colorado, Boulder, CO 80303, USA\\
$^6$ Laboratory for Atmospheric and Space Physics,
              3665 Discovery Drive, Boulder, CO 80303, USA\\
$^7$ Leibniz Institute for Astrophysics Potsdam, An der Sternwarte 16, 
	       14482 Potsdam, Germany\\
$^8$ ReSoLVE Centre of Excellence, Space Climate research unit, University of Oulu, P.O.Box 3000
FIN-90014 Oulu, Finland
}

\date{\today,~ $ $Revision: 1.199 $ $}

\begin{abstract}
Using a recently developed two-scale formalism to determine
the magnetic helicity spectrum \citep{BPS17}, 
we analyze synoptic vector magnetograms built with data from
the Vector Spectromagnetograph (VSM) instrument on the
\emph{Synoptic Optical Long-term Investigations of the Sun} (SOLIS) telescope
during January 2010--July 2016.
In contrast to an earlier study using only three Carrington rotations,
our analysis includes 74 synoptic Carrington rotation maps.
We recover here bihelical spectra at different phases of
solar cycle~24, where the net magnetic helicity in the majority of the
data is consistent with a large-scale dynamo with helical
turbulence operating in the Sun.
More than $20\%$ of the analyzed maps, however, show violations of the
expected sign rule.
\end{abstract}


\section{Introduction}

Magnetic helicity is a topological invariant of ideal
magnetohydrodynamics (MHD). It is a measure of complexity or
internal twist of the magnetic field structure and has a
geometrical interpretation in terms of linkage of magnetic
field lines \citep{BF84,AK92,Mof69}. Moreover, it is expected
to be nearly conserved even in non-ideal MHD systems with
large magnetic Reynolds number $\Rm$. This conservation law has been
recently tested in solar context involving magnetic reconnection
\citep{PVDD15}. Magnetic helicity thus plays a crucial role in the
evolution of magnetic fields and it can be an effective tracer of
the underlying mechanism responsible for the generation of
magnetic fields \citep{BS05a}.

There has been considerable interest in monitoring the magnetic
helicity of ARs, as this characterizes the complexity of the ARs
involved, and is therefore often linked to its ``eruptibility'', causing
solar flares and coronal mass ejection (CMEs);
see, e.g., \citet{NZZ03,Valori16}, and also \cite{Par17} who suggest a
better ``eruptivity proxy'' involving magnetic helicity.
Instead of being merely elements of the entire solar magnetic structure,
the ARs, and the magnetic helicity they carry, play an important
role for the global solar dynamo. 
The dynamo-generated large-scale field, by a mechanism still under
some debate, feeds the localized magnetic concentrations, leading
to the formation of ARs and sunspots.
The thereby formed ARs can contribute to migrate the
small-scale magnetic helicity, which is created as a by-product of the
helically-driven large-scale dynamo (LSD), away from the dynamo active
region, to prevent the quenching of the LSD \cite[see, e.g.,][and references therein]{LP03,BBS03,BS05a}. 

Although the origin of solar magnetism is yet
to be fully understood, it is commonly thought that the global
cyclic magnetic field of the Sun is generated and maintained by a
turbulent dynamo \citep{VZ72,Mof78,KR80,Ossen03,SIS06,Cha10}.
The solar dynamo is expected to involve an
$\alpha$ effect, which is a measure of the helicity of turbulence in the
convection zone caused by strong stratification and rotation
\citep{KR80}.
Numerical simulations have shown that a significant $\alpha$ effect is
indeed produced under these conditions in convective turbulence
\citep[e.g.][]{OSBR02,KKB09,WRTKKB18}.
It is known that the $\alpha$ effect produces a bihelical magnetic field
where the magnetic helicity at large and small scales have opposite
signs, and thus there is no net production of magnetic helicity in
the process \citep{See96,Ji99,BB03}.

\begin{figure}\begin{center}
\includegraphics[width=\columnwidth]{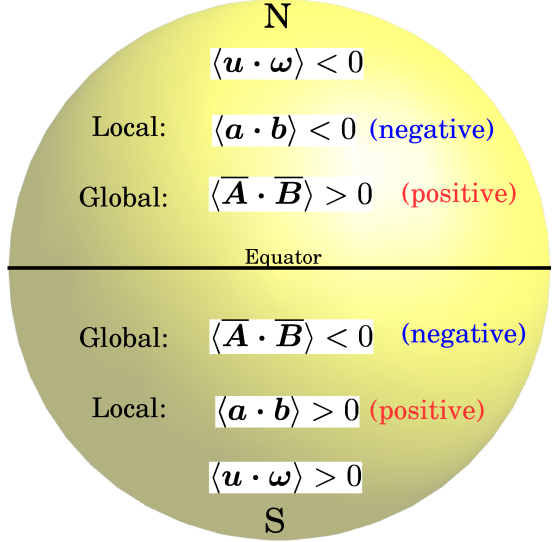}
\end{center}\caption[]{
A schematic diagram showing hemispheric sign rule of solar magnetic helicity
as expected from an $\alpha$-driven turbulent dynamo.
}\label{sign_rule}
\end{figure}

In the mean-field framework, the quantities, say, magnetic fields,
$\BB$, are expressed as a sum of mean, $\meanBB$, and fluctuating,
$\bb$, components, i.e., $\BB=\meanBB+\bb$, giving two contributions for the magnetic helicity:
${\cal H}_{\rm M}=\bra{\AAA\cdot\BB}=\bra{\meanAA\cdot\meanBB}
+\bra{\aaaa\cdot\bb}$, with $\AAA$ being the
vector potential defined from $\nab\times\AAA=\BB$
\citep{KR80,BS05a}.
Angle brackets $\bra{\,}$ denote the volume averages, while the
overbars indicate ensemble or longitudinal averages, satisfying the
Reynolds rules, for example $\bra{\aaaa}=\bf0$ and
$\bra{\meanAA\cdot\bb}=0$ \citep{KR80}.
This now helps us to summarize the expected hemispheric sign rule (HSR) of
solar magnetic helicity where the $\alpha$ effect changes sign across
the equator: the local (global) magnetic helicity is expected to be
negative (positive) in the northern hemisphere, and vice versa in the
southern hemisphere, as shown schematically in \Fig{sign_rule}.
The concept of scales is important in the present context where typical
extents of even the largest active regions (ARs) or the sunspots are
considered small, whereas scales of the order of the solar
radius are termed as large. 

The HSR is confirmed in a number of earlier works reporting measurements
of local as well as global magnetic net helicity at different phases
of the solar cycle (SC), using different techniques that often involve
determining the vector potential under a suitable gauge choice, see
e.g. the method of \cite{BBS03}.
Using this method on
data from the Michelson Doppler Imager
(MDI) on board the \emph{Solar and Heliospheric Observatory} (SOHO)
for SC~24 and SOLIS data for SC~23,
\cite{PP14} found 
that the global magnetic
helicity was indeed positive (negative) in the northern (southern)
hemisphere during SC~23 and SC~24, obeying thus the
HSR as shown in \Fig{sign_rule}.
The importance of such measurements in the solar context was discussed much
earlier \citep{See90} and many subsequent works, focussing mainly
on the ARs contributing thus to the local measurements of the helicity,
found that it is mostly negative (positive) in the north (south)---exactly
according to the expected sign rule
\citep[see, e.g.,][]{PCM95,BZAZ99,Liu14}.
\cite{LHS14} made an attempt to test the HSR using HMI data and found
that it was obeyed by nearly $75\%$ of all the ARs that they studied.

The dependence of magnetic helicity on the phase of the SC
has been explored to some extent.
\cite{BBS03} reported that the global magnetic helicity was
negative before the solar maximum and it turned positive afterwards,
i.e., they found evidence of a `wrong' sign during the rising phase
of the cycle. 
Similar results were obtained by \cite{PP14} for SC~23 and 24 from
MDI and SOLIS data.
Many studies also utilize the current helicity, 
$\mathcal{H}_{\rm C}$=$\left<\bm{J} \cdot \bm{B}\right>$,
where $\bm{J}=\mu_0^{-1} \nabla \times \bm{B}$ is the current density, 
as a proxy for the magnetic helicity, and argue that these quantities can
be used interchangeably \citep[e.g.][]{ZSPGXSK10}.
This holds strictly only for magnetic and current helicity spectra,
and only under isotropic conditions, which are in general
not met under solar conditions.
The study of \cite{ZSPGXSK10} showed that the current helicity
follows the equatorward propagation of magnetic dynamo wave traced through
the sunspots in the photosphere.  
While much of their analysis confirms the HSR, they do also notice
wrong signs of the helicity, mostly at the beginning and end
of the cycle, and interpreted this as due to penetration
of the activity wave into the other hemisphere.
Also employing the current helicity method on SOLIS data,
\cite{Gosain13}, however,  found no such violations
during early phase of SC~24.

It is only recently that, instead of computing net magnetic helicities
over a given domain,
methods for computing magnetic helicity distribution over different spatial
scales (spectrum) were developed.
These were first applied to local patches of photospheric magnetic
field measurements for a few ARs \citep{ZBS14,ZBS16}.
As the spectrum usually offers a much more detailed picture, it
allowed them to explain an earlier report on the net negative helicity
of an extremely complex AR~11515 which emerged in the southern
hemisphere \citep{Lim16}.
In order to also determine simultaneously the global spectrum,
\citet*[][hereafter BPS17]{BPS17} developed a two-scale formalism,
that allows us
to describe a fairly complex sign rule
of solar magnetic helicity, which depends on the position,
showing a systematic latitudinal dependence, as well as
scale.
They applied it to HMI data from three consecutive
Carrington rotations (CRs), 2161-2163, and found no evidence of
bihelical magnetic fields.

In the present work, we exploit the two-scale approach
to determine the solar magnetic helicity spectrum using SOLIS/VSM data
from 74~CRs covering more than six years of SC~24.
In \Sec{def} we review some basic definitions and outline the
two-scale approach.
In \Sec{da} we discuss the data and error estimation, and
in \Sec{res} we present 
the magnetic helicity spectra computed at various phases of SC~24.
We discuss the implications of our results and conclude in \Sec{dc}.

\section{Basic definitions and two-scale approach}
\label{def}

We first recall some fundamental aspects of the relevant physical
quantities and then briefly outline the two-scale approach recently
developed by BPS17 to determine a global spectrum of
magnetic helicity.
Let $B_i(\xx,t)$ denote the $i^{\rm th}$ component of the magnetic
field with $t$ being time, $\xx$
the position vector on the two-dimensional Cartesian surface, and
$i=(x,y,z)$.
The two-point correlation tensor of the total
magnetic field $\BB(\xx)$ is usually defined as
$M_{ij}(\xxi)=\bra{B_i(\xx)B_j(\xx+\xxi)}$ which is assumed to be
statistically independent of $\xx$ under homogeneous conditions,
where the brackets denote an ensemble average \citep{Bat53,Mof78}.
We omit specifying explicitly the temporal dependencies from now on.
The spectrum of magnetic energy, $E_{\rm M}(k)$, is then given by
\EQ
2E_{\rm M}(k)=\int \delta_{ij}\hat{M}_{ij}(\kk)\,k\,\dd\Omega\,,
\label{emspec}
\EN
where
\EQ
\hat{M}_{ij}(\kk)=\int M_{ij}(\xx)\,e^{-\ii\kk\cdot\xx}\,
\dd^2 x/(2\pi)^2
\label{ft}
\EN
is the two-dimensional Fourier transform of $M_{ij}$ and the
wavevector $\kk$ denotes the conjugate variable to $\xx$
and $E_{\rm M}(k)$ is measured in $\G^2\cm$ rather than $\erg\cm^{-2}$.
In two dimensions, $\int \dd\Omega=2\pi$ is the circumference of a
unit circle and the integral in \Eq{emspec} is performed over shells
in wavenumber space.
The scaled magnetic helicity spectrum, $kH_{\rm M}(k)$, which has the
same dimensions as that of $E_{\rm M}(k)$, is similarly defined as
\EQ
kH_{\rm M}(k)=\int\ii\hat{k}_i\epsilon_{ijk}\hat{M}_{jk}(\kk)\,k\,
\dd\Omega\,,
\label{kHkspec}
\EN
where $\hat{k}_i=k_i/|\kk|$ is the unit vector of $\kk$ and $k=|\kk|$
is its modulus with $k^2=k_x^2+k_y^2$.
Thus, the spectra of magnetic energy and helicity can be determined
from the two-point correlation function using
\Eqs{emspec}{kHkspec} where the former is given by the trace of the
Fourier transform of $M_{ij}$ resulting in a positive-definite scalar
quantity, $E_{\rm M}$, whereas the latter is defined by the skew-symmetric
part of $\hat{M}_{jk}$ giving a pseudo-scalar quantity, $H_{\rm M}$,
which can take both positive and negative values \citep{Bat53,Mof78,BS05a}.

As discussed in BPS17, relaxing the assumption of homogeneity
allows us to determine the spectra as a function of slowly varying
coordinate denoted by, say, $\XX$. This is particularly relevant for the
Sun where we expect opposite signs of helicities in the northern and
southern hemispheres, while assuming statistically similar conditions
at all longitudes. Below we describe such a procedure to determine
spectra that involve a double-Fourier transform.

\subsection{The two-scale approach}
\label{tsa}

Under non-homogeneous conditions, the two-point correlation function,
$M_{ij}(\xx',\xx'')=\bra{B_i(\xx')B_j(\xx'')}$ takes the form
\citep{RS75}:
\EQ
M_{ij}(\XX,\xx)=\bra{B_i(\XX+\half\xx)\,B_j(\XX-\half\xx)},
\label{mijts}
\EN
where $\XX=(\xx'+\xx'')/2$ is the mean or slowly varying coordinate
and $\xx=\xx'-\xx''$, called the relative coordinate, is the
distance between the two points around $\XX$. 
Fourier transforming \Eq{mijts} first over $\xx$, and then over $\XX$
after assuming locally isotropic conditions, one obtains the
following simple expression for the doubly Fourier transformed
two-point correlation function (BPS17):
\EQ
\tilde{M}_{ij}(\KK,\kk)=\bra{\hatB_i(\kk+\half\KK) \,
\hatB_j^\ast(\kk-\half\KK)}.
\label{mijKk}
\EN
Here the wavevectors $\KK$ and $\kk$ denote the conjugate variables
to $\XX$ and $\xx$, respectively.
Analogously to \Eqs{emspec}{kHkspec}, the $\KK$-dependent magnetic
energy and helicity spectra are thus determined from (BPS17):
\EQ
2\tilde{E}_{\rm M}(\KK,k)=\int \delta_{ij}\tilde{M}_{ij}(\KK,\kk)
\,k\,\dd\Omega,
\label{2EKk}
\EN
\EQ
k\tilde{H}_{\rm M}(\KK,k)=\int \ii\hat{k}_i\epsilon_{ijk}
\tilde{M}_{jk}(\KK,\kk)
\,k\,\dd\Omega.
\label{kHKk}
\EN
The spectrum of magnetic helicity with a slow variation in
the $z$ direction is proportional to $\sin K_Z Z$ and is given by
$\KK=(0,0,K_Z)$, where $K_Z=2\pi/L$ and $z=Z$ are used interchangeably.

Unlike $H_{\rm M}(\XX,k)$, which is real, $\tilde{H}_{\rm M}(\KK,k)$
is complex.
The quantity of interest depends on the spatial profile of the
background helicity.
We are here concerned with helicity profiles
proportional $\sin K_0 Z$ with an equator at $Z=0$.
Its Fourier transform is $-\half\ii\delta(K_Z-K_0)$.
We will therefore plot the {\em negative imaginary part} of
$\tilde{H}_{\rm M}(\KK,k)$, which reflects the sign of
magnetic helicity in the northern hemisphere.
The total magnetic energy ${\cal E}_{\rm M}$ and helicity ${\cal H}_{\rm M}(K_0)$
are defined as,
\EQA
\label{Etot}
{\cal E}_{\rm M}&=&\int_0^\infty \tilde{E}_{\rm M}(0,k)\,\dd k,\\
\label{Htot}
{\cal H}_{\rm M}(K_0)&=&\int_0^\infty \tilde{H}_{\rm M}(K_0,k)\,\dd k,
\ENA
which will be used in \Sec{res}.

\section{Data analysis}
\label{da}

We analyze synoptic vector magnetograms from 74 CRs
of the Sun where we determine the magnetic energy and helicity spectra
either for each CR, or by sometimes first combining the synoptic vector
magnetograms from three successive CRs.
The data is based on measurements from the Vector SpectroMagnetograph
(VSM) instrument on the Synoptic Optical Long-term Investigations
of the Sun (SOLIS) project
\citep{Keller_2003,Balasubramaniam_2011}.
SOLIS/VSM observes Fe I 630.15 and 630.25 nm spectral lines,
with a spatial sampling of 1.14'' per pixel and 2048 $\times$ 2048 pixels field of view.
The line profiles of Stokes $Q$, $U$, $V$, and $I$ are derived using the Very Fast Inversion of Stokes
Vector code \citep{Borrero.etal2011} on the Fe I 630.25~nm line,
which includes the magnetic field filling factor.
The 180$^\circ$ ambiguity in the transverse field direction is solved
using the Very Fast Disambiguation Method \citep{Rudenko_2014}.
Synoptic maps of the three vector components of the photospheric magnetic field are constructed from daily full-disk magnetograms. 
We use the 180 by 360 pixel maps of the photospheric vector magnetic
field, where each pixel gives the observed full vector magnetic field
$\BB\equiv(B_r,B_\theta,B_\phi)$, with $r$, $\theta$, and $\phi$ corresponding
to the radius, colatitude and longitude, respectively.
The field is mapped onto the $(\phi,\mu)$ plane with $\mu=\cos\theta$,
allowing us to adopt a right-handed Cartesian analysis by substituting:
\EQ
(\phi,\mu)\to(y,z),\quad (B_r,B_\phi,-B_\theta)\to(B_x,B_y,B_z).
\EN
In the definitions of quantities of interest, notably
$k\tilde{H}_{\rm M}(\KK,k)$, given in \Sec{tsa}, we consider
a fixed wavevector $\KK=(0,K_0)$ where $K_0=\pi/R_\odot$ with
$R_\odot$ being the solar radius, and determine this as function
of $k$.
Thus we assume that the background helicity does not have any
systematic modulation in longitude. For the energy spectrum,
we consider no modulation, as usual, and determine
$\tilde{E}_{\rm M}(0,k)$ versus $k$.

We consider all CR numbers between 2093 -- 2178, except CRs 2099, 2107,
2127, 2139, 2152 -- 2155, 2163, 2164, 2166, and 2167.
These rotations suffer from poor data coverage and therefore depict
obvious outliers. 
Our analysis thus covers a period from 2010-01-30 to 2016-07-03 of the current
SC~24, which reached its maximum during the middle of the year 2014.

The wavelength-dependent scatter of the spectra can be considered as
a measure of the error introduced by the temporal evolution of the 
synoptic maps and, to a smaller extent, stochastic errors in the 
measurement of the magnetic field vector.
We define the root-mean-squared (rms) error $\sigma_{\cal P}(k)$ associated
with the spectrum ${\cal P}(k)$ as
\EQ
\sigma_{\cal P}(k) = \sqrt{\bigg\langle \Bigl(
{\cal P}(k)-\bra{{\cal P}(k)}_{\rm CR} \Bigr)^2
\bigg\rangle}\,,
\label{err}
\EN
where $\bra{\;\;}_{\rm CR}$ denotes the average over CRs~2148--2151,
which corresponds to the period of maximum solar activity. The 
statistical error adopted here is expected to be largest at this phase
and therefore it is safer to read the spectra even from other epochs
in the light of over-estimated errors being shown.

This error, however, does not contain the uncertainty of the magnetic
field measurement itself.
Noise in spectral line observations, uncertainties and simplifications
in inversion method  (like assumption of a Milne-Eddington type atmosphere)
and possible errors in disambiguation method introduce uncertainties.
It is virtually impossible to reliably quantify this error
\cite[see, e.g.,][]{JMB14}. 
Moreover, synoptic map is constructed from consecutive observations
over solar rotation, it is not a snapshot. 
Averaging over large number of pixels observed within few days
improves the signal-to-noise ratio, but makes the variable small-scale
magnetic features less reliable.

\begin{figure}\begin{center}
\includegraphics[width=\columnwidth]{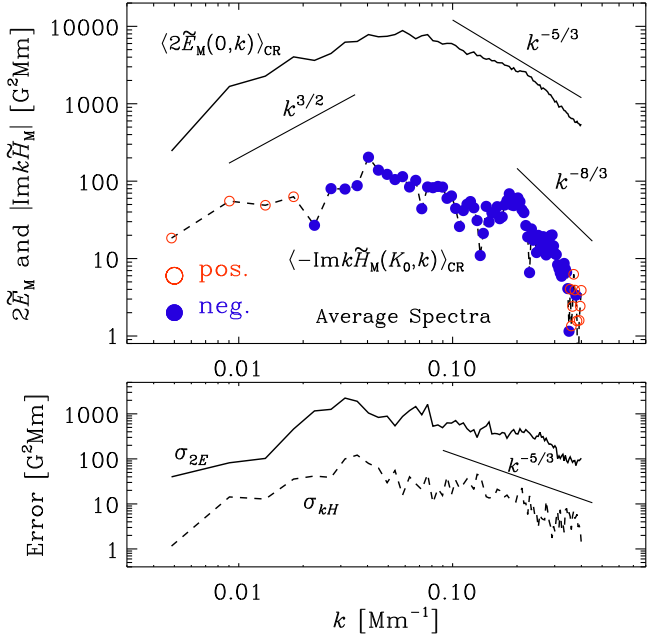}
\end{center}\caption[]{
Top: magnetic energy (solid) and helicity (dashed; circles) spectra
obtained after averaging spectra from CRs~2148-2151.
Sign convention adopted here corresponds to the sign of magnetic helicity
in the northern hemisphere; open red (filled blue) circles
denote positive (negative) signs for the magnetic helicity.
Bottom: errors on $\bra{2\tilde{E}_{\rm M}}_{\rm CR}$ and
$\bra{-\Imag k\tilde{H}_{\rm M}}_{\rm CR}$, as determined from \Eq{err},
are shown by solid and dashed lines, respectively.
}\label{kavgd_specmk}
\end{figure}

\section{Results}
\label{res}

\subsection{Average spectrum during the maximum of SC~24} \label{max}

First we show, in \Fig{kavgd_specmk}, the spectra of magnetic helicity
and energy as a function of $k$,
obtained after averaging over individual spectra
from CRs 2148 -- 2151 which correspond to the phase when the Sun
was most active during SC~24 in terms of sunspot number.
As noted before, the relevant quantity here is
$-\Imag\, k\tilde{H}_{\rm M}(K_0,k)$, which has the sign convention
corresponding to the northern hemisphere. Remarkably, the averaged
spectrum of solar magnetic helicity, denoted by
$\bra{-\Imag k\tilde{H}_{\rm M}(K_0,k)}_{\rm CR}$,
as shown in the top panel of \Fig{kavgd_specmk},
clearly reveals a bihelical signature,
with positive (negative) helicity at small (large) wavenumbers, exactly
as would be expected from an $\alpha$ effect-driven solar dynamo
\citep{BB03,YB03}; see \Fig{sign_rule} for expected HSR.
However, the power at small $k$ is rather weak.
The $k$-dependent errors, $\sigma_{2E}$ and $\sigma_{kH}$,
estimated according to \Eq{err}, are shown in the bottom panel of
\Fig{kavgd_specmk}.
The retrieved magnetic helicity spectrum during the maximum phase
of SC~24 is significant at all $k$ when contrasted to its error.

In the context of the solar dynamo, a distinction between large and
small scales can be made based on the wavenumber $k$, where the magnetic
helicity changes sign, i.e., at $k\approx 0.02\Mm^{-1}$, corresponding
to a scale ${\cal L}\approx 315\Mm$ from \Fig{kavgd_specmk}.
Typical scales associated with ARs of $\sim30\Mm$
are therefore considered ``small'', whereas the large scales are comparable
to the solar radius.
We also note that such a distinction is not always possible as the
spectrum shows significant variation between different epochs.
Nevertheless, it gives us a perspective on the relevant scales
involved in the underlying dynamo mechanism.

We recall that an energy spectrum proportional to $k^2$ (Saffman spectrum)
means that the large-scale field is random.
However, only the steeper $k^4$ Batchelor spectrum would imply that the
largest scales are not causally related to the smaller ones \citep{DC03}.
All the spectra retrieved in this study show shallower powerlaws
at the smallest wavenumbers, implying 
a causal connection between the large and small scales.
As discussed further below, it is also possible that we see evidence
of Kazantsev scaling with a $k^{3/2}$ subrange at $k\Mm<0.03$,
which would be indicative of a small-scale dynamo (SSD) \citep{Kaz68}.
Indeed, astrophysical dynamos operating at high magnetic Reynolds
numbers are expected to exhibit a unified version of dynamo action
that combines elements of both small-scale and large-scale dynamos
\citep{Sub99,SB14,BSB16}.
On the other hand, the bihelical magnetic field expected from an
$\alpha$ effect is usually expected to imply an actual increase in
magnetic power at small $k$; see Figure~3 of \cite{B01}.
All the spectra computed here show less power at small wavenumbers
than in the large ones; this could be a manifestation of the SSD
dominating the LSD near the surface. 
dominating the LSD near the surface \citep{TrujilloBueno_etal04}.

At the largest values of $k$, one also sees occasional data points with
a reversed sign, but the power is again low.
The measured power is dropping below its estimated error
at wavenumbers where the sign change occurs, so this cannot be
regarded as a reliable finding.
Nevertheless,
the occurrence of mixed signs as such is not surprising given the turbulent
nature of the underlying magnetic field and has been seen before
(BPS17). However, compared with the usual one-scale approach used in
\cite{ZBS14,ZBS16}, these mixed signs are surprisingly rare.

\begin{figure}\begin{center}
\includegraphics[width=\columnwidth]{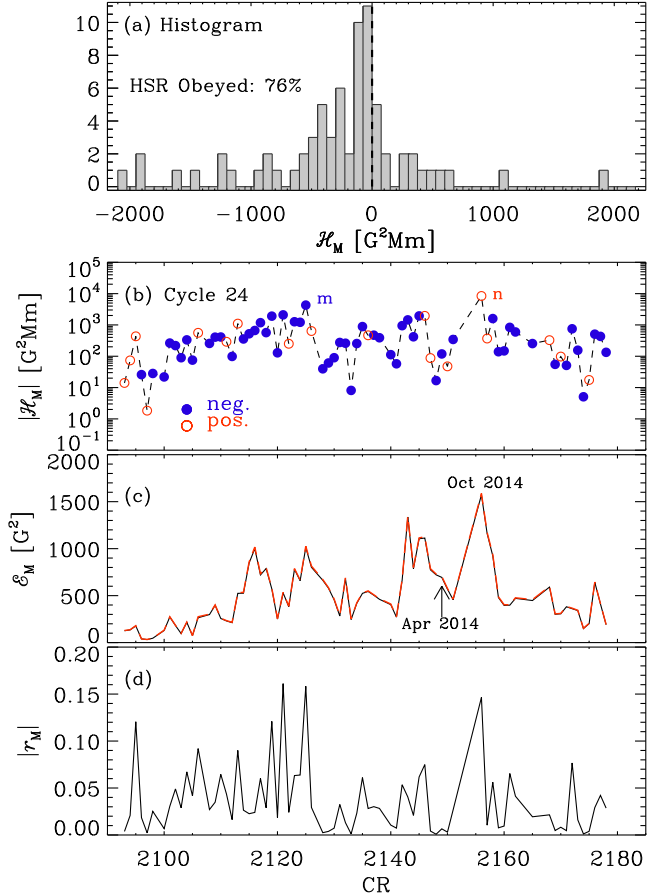}
\end{center}\caption[]{
(a) Histogram of ${\cal H}_{\rm M}$,
temporal evolution of (b) $|{\cal H}_{\rm M}|$, (c) ${\cal E}_{\rm M}$,
and (d) $|r_{\rm M}|$;
see \Eqs{Etot}{Htot}.
The blue, filled (red, open) circles in panel (b) denote, as usual,
negative (positive)
values with sign convention corresponding to the northern hemisphere.
The two largest values of $|{\cal H}_{\rm M}|$ marked by letters `m'
and `n' are ignored in the histogram.
Solid, black (dashed, red) curves in (c) correspond to ${\cal E}_{\rm M}$
from \Eq{Etot} and directly from synoptic maps, respectively.
}\label{tev_EHl}
\end{figure}

\subsection{HSR statistics}\label{HSR}

Before discussing more spectra from a few individual CRs at different
phases of SC~24, we next look at the
temporal evolution of the total integrated magnetic energy ${\cal E}_{\rm M}$
and helicity ${\cal H}_{\rm M}$, defined in \Eqs{Etot}{Htot}, respectively.
These are determined by first obtaining the corresponding spectra for each
CR using \Eqs{2EKk}{kHKk} and then computing the $k$ integral.
The integrated magnetic helicity is shown in \Fig{tev_EHl}(a) and (b),
which reveals that it is generally
small, as might be expected for bihelical magnetic fields leading
to significant cancellations of opposite helicities at large and small
scales.
As is evident from the histogram presented in \Fig{tev_EHl}(a), the 
most common values are around a few tens of $\G^2\Mm$, while the distribution
also develops wide wings with values of the order of one or two thousand $\G^2\Mm$, but such
events are relatively rare.
They can be associated with
complex ARs dominating the spectrum with significant intrinsic
magnetic helicity. We discuss some examples later in this paper.

The median of the distribution is clearly negative, as can also be seen 
in the dominance of blue circles in \Fig{tev_EHl}(b).
This is due to the large-scale contributions, giving a positive signal
in the northern hemisphere if HSR is obeyed, being
sub-dominant to the negative helicity carried by
the ARs. Therefore, positive values of this quantity can indicate either
an occasionally dominating positive large-scale contribution, or a non-HSR 
obeying positive helicity at the smaller scales. The former happens only during
the early declining phase of SC~24, when magnetic energy and helicity
obtain maxima, as will be 
discussed in detail in \Sec{edp}.  
Therefore, a negative (positive) sign of the integrated magnetic
helicity can be regarded as a good proxy obedience (violation) of HSR.

Of all the 74 CRs analyzed, 76\% exhibit negative integrated
magnetic helicity, and are therefore judged to obey the HSR.
Note again that we follow here the sign convention of northern hemisphere,
which may be inferred from \Fig{sign_rule}.
The likelihood to obey HSR is increased during the
ascending phase of the SC, while it is decreased during the first
few CRs of SC~24.
The integrated magnetic energy, shown in \Fig{tev_EHl}(c), attains
a maximum in Oct 2014, when also very large magnitudes of
magnetic helicity are seen. This is roughly 6 months 
later than the maximum of SC~24 obtained from sunspot
numbers (April 2014). Already before the maximum energy and
helicity are reached, the sign of the integrated magnetic helicity 
becomes ill-defined, the reasons behind this being discussed in \Sec{erp}, 
and this behavior continues during the declining phase; see \Sec{edp}. 

It is useful to have some estimate of integral scale of turbulence
$\ell_{\rm M}$, which is defined as,
\EQ
\ell_{\rm M}=\left.\int_0^\infty k^{-1} \tilde{E}_{\rm M}(0,k)\,\dd k\,
\right/\int_0^\infty\tilde{E}_{\rm M}(0,k)\,\dd k\,.
\label{ell}
\EN
But due to the low resolution SOLIS data being considered here,
resulting in less clear power-law scalings in the magnetic energy
spectra for high values of $k$, we do not find reliable estimates of $\ell_{\rm M}$.
These estimates are expected to be much better from high resolution
HMI data used by BPS17 who found that $\ell_{\rm M}\approx 20\Mm$.
We know from the realizability condition \citep{Mof78,KTBN13}, i.e.,
$|{\cal H}_{\rm M}|/2{\cal E}_{\rm M}\leq\ell_{\rm M}$, that
the magnetic energy of helical fields is bounded from below and therefore
the absolute value of the quantity $r_{\rm M}={\cal H}_{\rm M}/
2\ell_{\rm M}{\cal E}_{\rm M}$ cannot exceed unity.
In \Figp{tev_EHl}{d}, we show the evolution of
$|r_{\rm M}|$ and note that the
realizability condition is obeyed at all times, with $|r_{\rm M}|$
being always below 0.2.
This is similar to what was obtained in BPS17.

\begin{figure}\begin{center}
\includegraphics[width=\columnwidth]{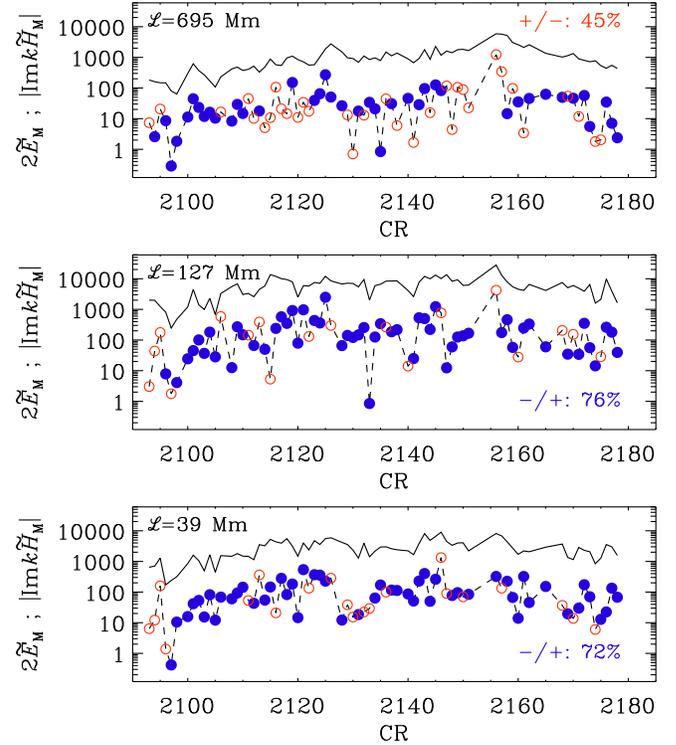}
\end{center}\caption[]{
Temporal evolution of spectral magnetic energy and helicity
at $k\approx0.01\Mm^{-1}$ (top), $0.05\Mm^{-1}$ (middle) and
$0.16\Mm^{-1}$ (bottom), where corresponding length scale
${\cal L}=2\pi/k$ is displayed in panels;
``$+/-$'' (``$-/+$'') denotes the percentage fraction of positive (negative)
sign of the magnetic helicity at chosen $k$.
}\label{ls_ss_tev}
\end{figure}

It is also important to check how well the HSR proxy, based on the total
magnetic helicity, works by inspecting
how the sign of magnetic helicity changes
at a few selected values of $k$.
With the sign convention of the northern hemisphere as before,
we show in \Fig{ls_ss_tev} the temporal evolution of
$2\tilde{E}_{\rm M}$ and $-\Imag k\tilde{H}_{\rm M}$ at three fixed
values of $k$. Again, most of the analyzed data reveal that the expected
HSR is obeyed, as may be seen from the bottom two panels corresponding
to intermediate and small scales, dominated mostly by the ARs, which
are expected to carry a net negative helicity in the north.
However, the top panel corresponding to large scales shows a
larger fraction of CRs violating the HSR. The absolute values of
the helicity are indeed much smaller at these scales and better
estimates are therefore needed to reliably determine the sign of
magnetic helicity at large scales.

\begin{figure}\begin{center}
\includegraphics[width=\columnwidth]{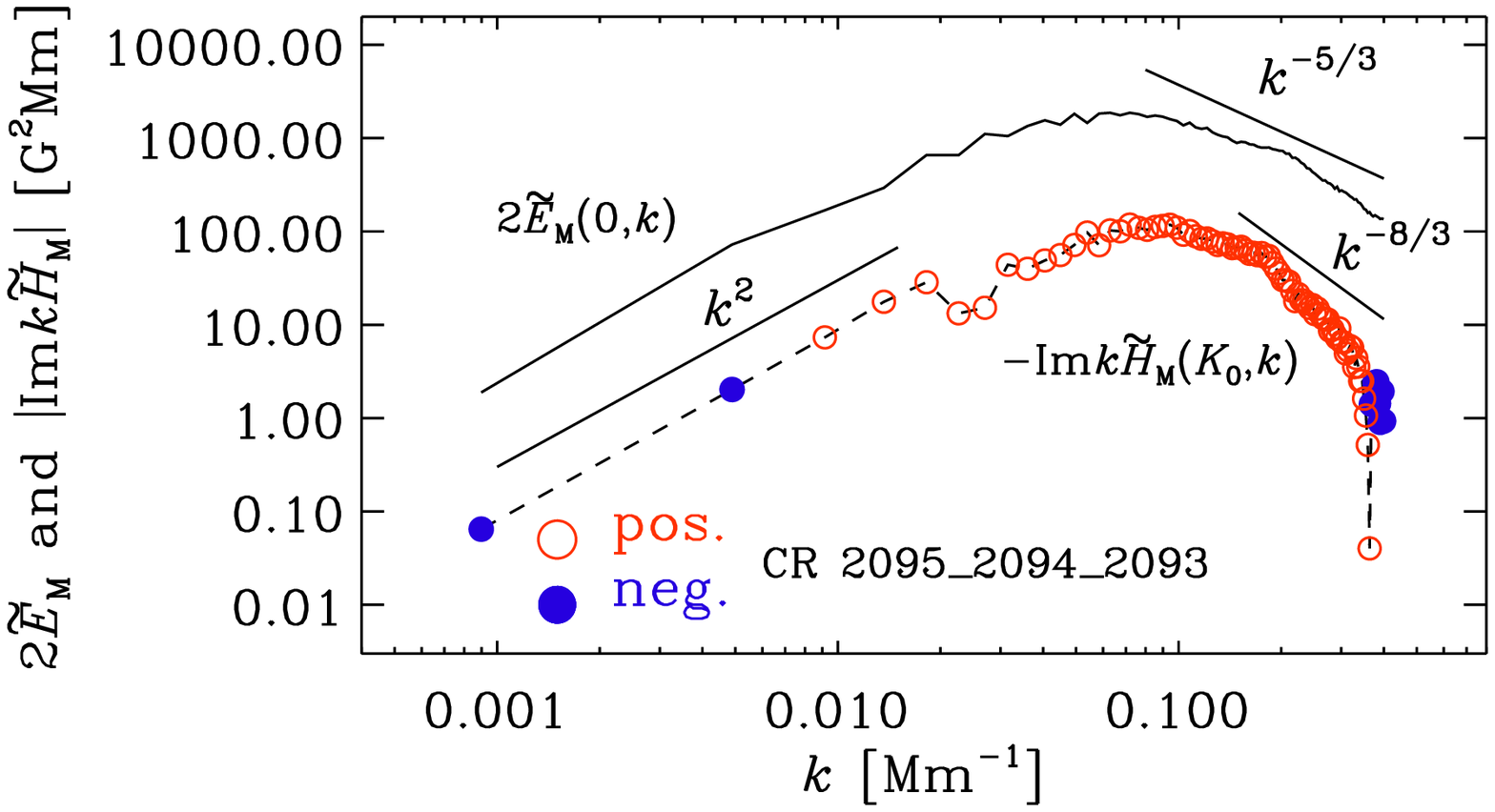}
\includegraphics[width=\columnwidth]{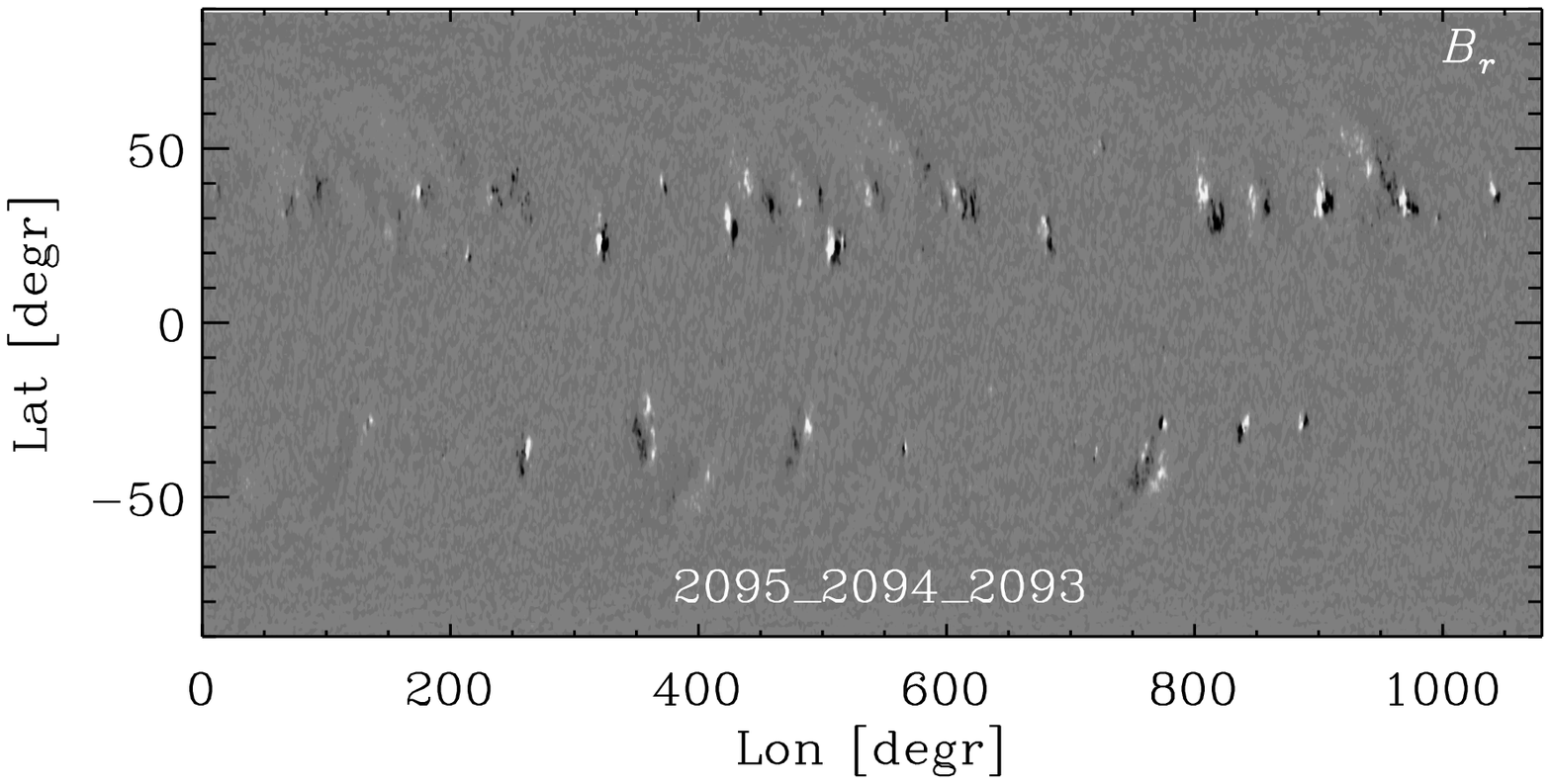}
\end{center}\caption[]{
Top: magnetic helicity and energy spectra from the interval spanning
CR~2093--2095. Spectra are determined after stitching together
data from these CRs.
Bottom: synoptic chart of radial component of the magnetic field, $B_r$,
covering the same time span. The sign convention adopted here corresponds
to the sign of magnetic helicity in the northern hemisphere;
open (filled) circles denote positive (negative) signs for the
magnetic helicity.
}\label{2095-2093}
\end{figure}

\subsection{Early rising phase of SC~24}\label{erp}

We now show in \Fig{2095-2093} the spectra of magnetic helicity and
energy that are obtained after stitching together data from three consecutive
CRs~2093--2095, which correspond to the early rising phase
of SC~24 covering a period from 2010-01-30 to 2010-04-21.
The corresponding synoptic maps of only the radial component of
the magnetic field, $B_r$, is also shown in \Fig{2095-2093}.
Note that with $\phi$ being longitude,
the range $0\degr\leq\phi<360\degr$ refers to CR~2095,
$360\degr\leq\phi<720\degr$ refers to CR~2094, and
$720\degr\leq\phi\leq1080\degr$ refers to CR~2093.

The magnetic energy and helicity peak at smaller scales, approximately
at 0.07 and $0.09\Mm^{-1}$, than during the maximum phase, but obtain
similar magnitudes than during the maximum at their peak values. The
large-scale powers are very weak, and fall below the estimated errors.
The spectral scaling is steep, close to Saffman spectrum with $k^2$, 
indicating random large-scale fields, but due to the weak signal
large uncertainty is related to this value.

Although the magnetic fields are clearly bihelical, the signs of
magnetic helicity at small and large $k$ are exactly opposite of
what we expect from a simplistic turbulent dynamo model. 
\cite{CCN04}, however, discussed a 
different (Babcock-Leighton type) dynamo model that can predict such violations of HSR 
during the early phase of the cycle. These arise due to the flux tubes of the new
cycle emerging in regions where poloidal fields from the previous cycle, possessing 
helicity of a wrong sign, still persist.
Turbulent dynamo models can also produce similar sign reversals when magnetic
helicity conservation law is used to constrain the model \citep{Pipin2013}.

In comparison to other observational results, current helicity based proxies
indicate such reversals for SC~22 \citep{BAZ00,ZSPGXSK10}, while 
not for SC~24 \citep{Gosain13}, and the results for SC~23 remain contradictory,
\cite{PCL01} against and \cite{ZSPGXSK10} in favor of reversals.
In contrast, \cite{PP14}
results computing the global magnetic helicity using azimuthally averaged mean
magnetic field, indicate a sign reversal at the large scales in the early phases of SC~24.

From the magnetogram showing the radial magnetic field $B_r$
in \Fig{2095-2093}, we see that most ARs are located
at higher latitudes, as expected if the ARs followed the butterfly diagram 
typical for early rising phase, and therefore we do not expect significant
``leakage'' of magnetic helicity of opposite sign through the equator.
It appears more likely that the ARs are
intrinsically twisted in an opposite sense and
dominate the
magnetic helicity spectrum in \Fig{2095-2093} with small-scale positive
helicity in the northern hemisphere, showing a maximum at scales around
$60\Mm$.

\begin{figure}\begin{center}
\includegraphics[width=\columnwidth]{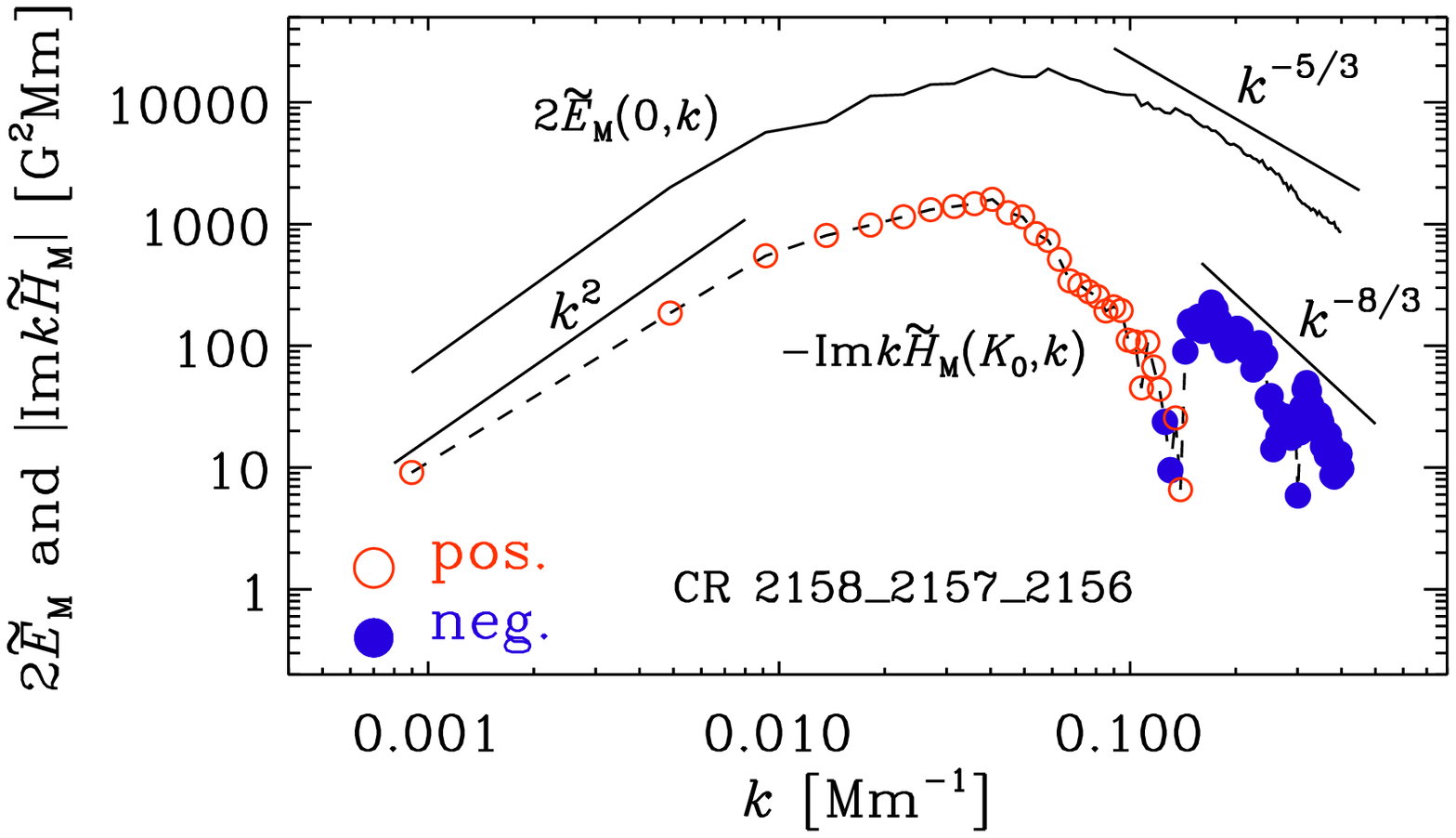}
\includegraphics[width=\columnwidth]{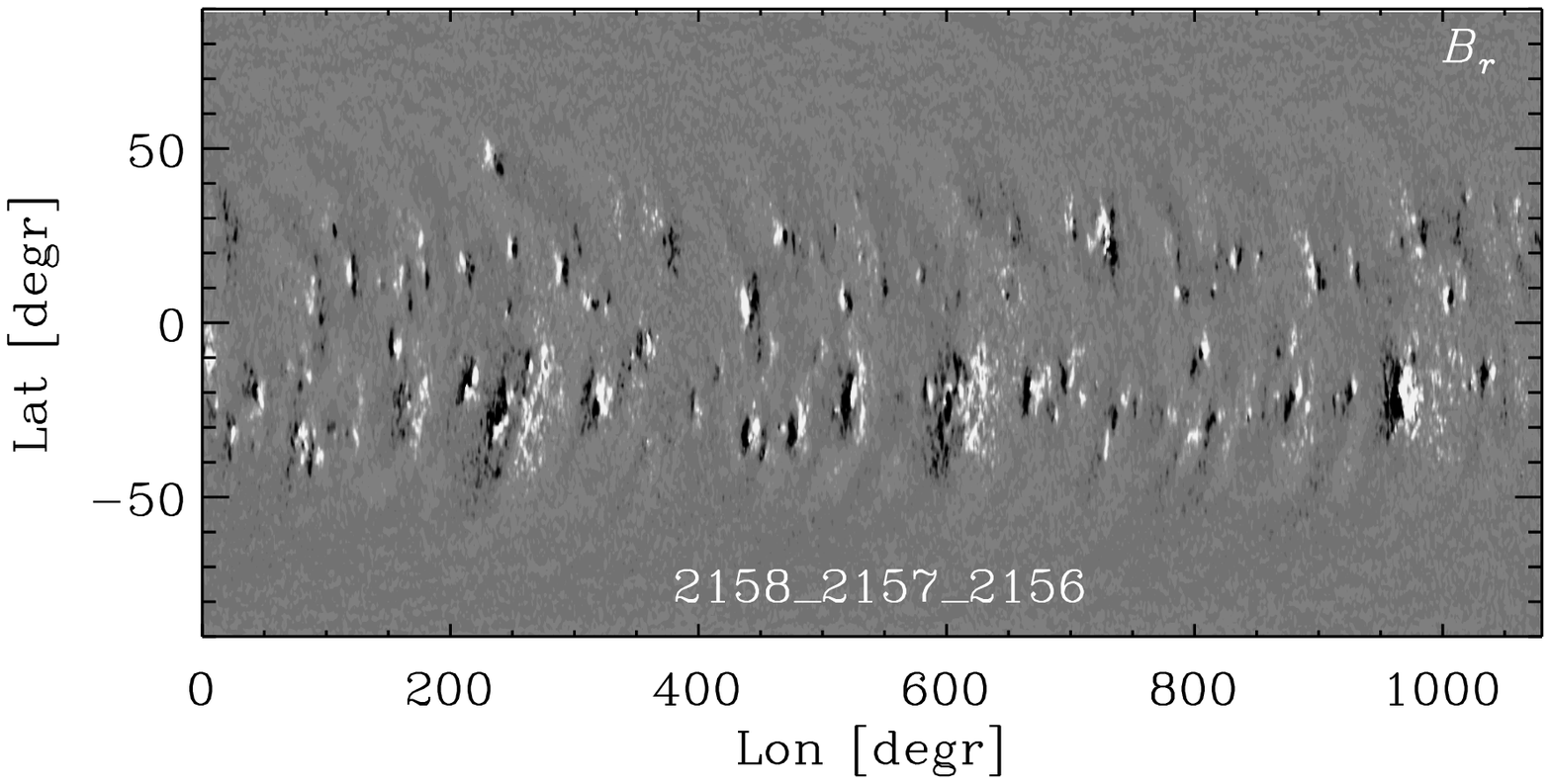}
\end{center}\caption[]{
Same as \Fig{2095-2093} but for the interval spanning CR~2156--2158.
}\label{2158-2156}
\end{figure}

\subsection{Early declining phase of SC~24} \label{edp}

Similarly, in \Fig{2158-2156} we show the spectra and
the magnetogram from the interval spanning CR~2156--2158,
which corresponds to the period from 2014-10-15 to 2015-01-03,
i.e., just after the maximum of SC~24.
While SC~24 reached its maximum in terms
of the actual number of sunspots in April 2014, the energy and also the
helicity show a peak during October 2014 corresponding to CR~2156; see
\Fig{tev_EHl}. The magnetic helicity spectrum now once again
shows a bihelical signature with signs at small and large $k$ being
compatible with the HSR based on an $\alpha$-driven solar dynamo.
This is qualitatively similar to the averaged helicity spectrum shown in
\Fig{kavgd_specmk}, but we also note an important difference.
Here we find a positive sign for the peak value of
$-\Imag k\tilde{H}_{\rm M}(K_0,k)$ at $k_{\rm peak}\approx0.04\Mm^{-1}$,
i.e., scales around ${\cal L}_{\rm peak}\approx160\Mm$, with the spectrum
turning negative for $k>0.1\Mm^{-1}$, i.e, at scales smaller than $60\Mm$.
Hence, while the magnetic helicity spectrum at the largest and
very smallest scales remains largely unaltered, in the mid-range
scales though, where usually the ARs dominate with strongly negative
helicity, we observe strong reversed, i.e., positive magnetic helicities.
As a result, the total solar magnetic helicity ${\cal H}_{\rm M}$ during this
period is positive in the northern hemisphere (marked by the letter `n' in
the inset of \Figp{tev_EHl}{b}), thus appearing to violate the HSR,
defined based on the sign of the total magnetic helicity.

However, a closer look presents a much richer
picture accessible only through a spectrum such as the ones being
explored here. Comparing the integral scale of turbulence
$\ell_{\rm M}\approx20\Mm$ as noted below \Eq{ell} to ${\cal L}_{\rm peak}$
determined above gives a scale separation
$\zeta={\cal L}_{\rm peak}/\ell_{\rm M}\approx8$.
Assuming this to be sufficient for distinguishing between the large and
small scales, we let, in this case, ${\cal L}_{\rm peak}$ to represent
the `large' scale. Then the helicity spectrum in \Fig{2158-2156} is
reminiscent of a classic picture due to an LSD where
the spectra have a peak at scales that are considerably larger than the
turbulent scales.
Interestingly enough, \cite{SW15} found that the Sun's large-scale magnetic
field was rejuvenated exactly during this period. This is further supported
by our inference that the power spectra are dominated by the LSD
during CR~2156-2158, thus resulting in positive ${\cal H}_{\rm M}$
in the northern hemisphere without, it seems, violating the HSR.

\begin{figure}\begin{center}
\includegraphics[width=\columnwidth]{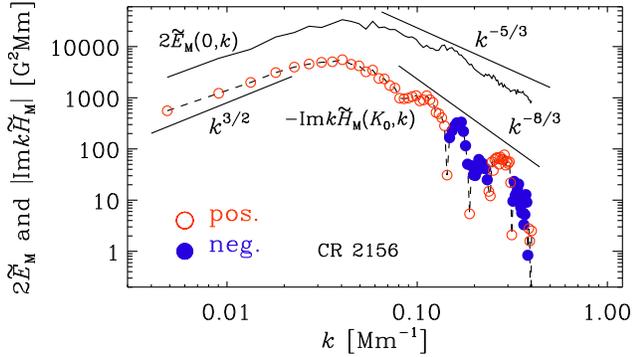}
\end{center}\caption[]{
Same as \Fig{kavgd_specmk} but for CR~2156 which corresponds to the
time when both magnetic energy and helicity reach a maximum in \Fig{tev_EHl}.
}\label{max2156_spec}
\end{figure}

To examine in more detail the epoch when both total magnetic energy and
helicity maximize, we zoom into CR~2156 and compute the spectra for it alone
(hence the shorter $k$-range)
in \Fig{max2156_spec}, which, except for showing sign fluctuations in
$-\Imag k\tilde{H}_{\rm M}$ at large $k$,
look otherwise similar to \Fig{2158-2156}.
The small-scale sign fluctuations might also be caused by many complex
ARs, such as 12192, 12205, 12209, 12241, 12242 etc, being, at times,
of the $\delta$-type, that could carry intrinsic helicities that are not necessarily
always according to the sign rule.
For SC~23, the number of complex ARs was found to
decline slower than the total number of ARs, due to which
their relative fraction was observed to be higher during the declining phase
\citep{JN16}, which lends support to this scenario.
We note, in addition, that the power in the largest scales is significantly
enhanced during this CR, an indication of enhanced LSD during this epoch.

Intriguingly, the magnetic energy spectrum shows a Kazantsev scaling of
$k^{3/2}$, which is predicted for the SSD,
albeit for the sub-inertial range.
Here this scaling is, rather unexpectedly, seen at the large scales. 
As we elaborate in \Sec{dc}, these results are suggestive of both LSD and SSD being operative
simultaneously in the Sun, with the $k^{3/2}$ scaling due to the SSD and bihelicity
of fields due to the LSD from an $\alpha$-effect.

\begin{figure}\begin{center}
\includegraphics[width=\columnwidth]{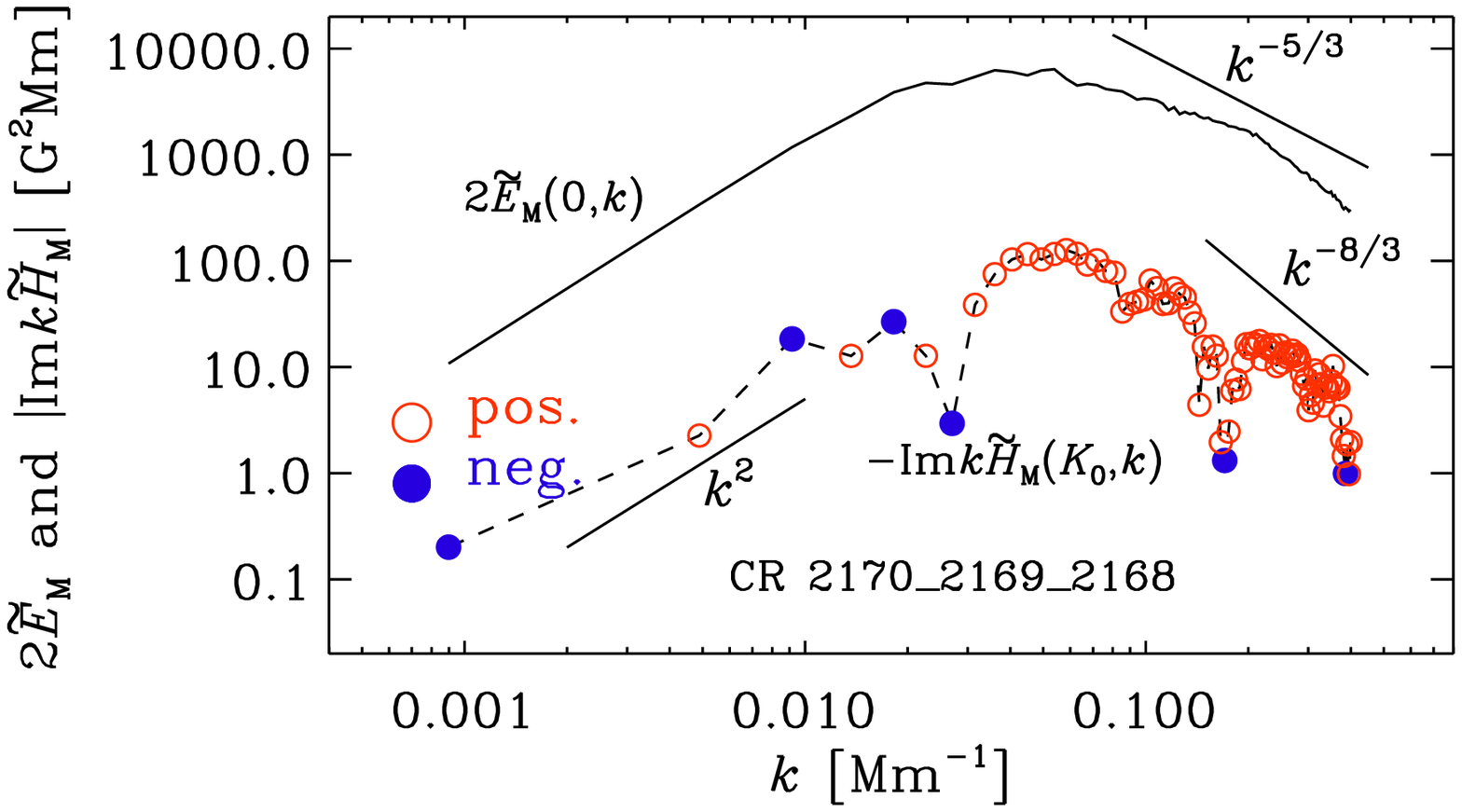}
\includegraphics[width=\columnwidth]{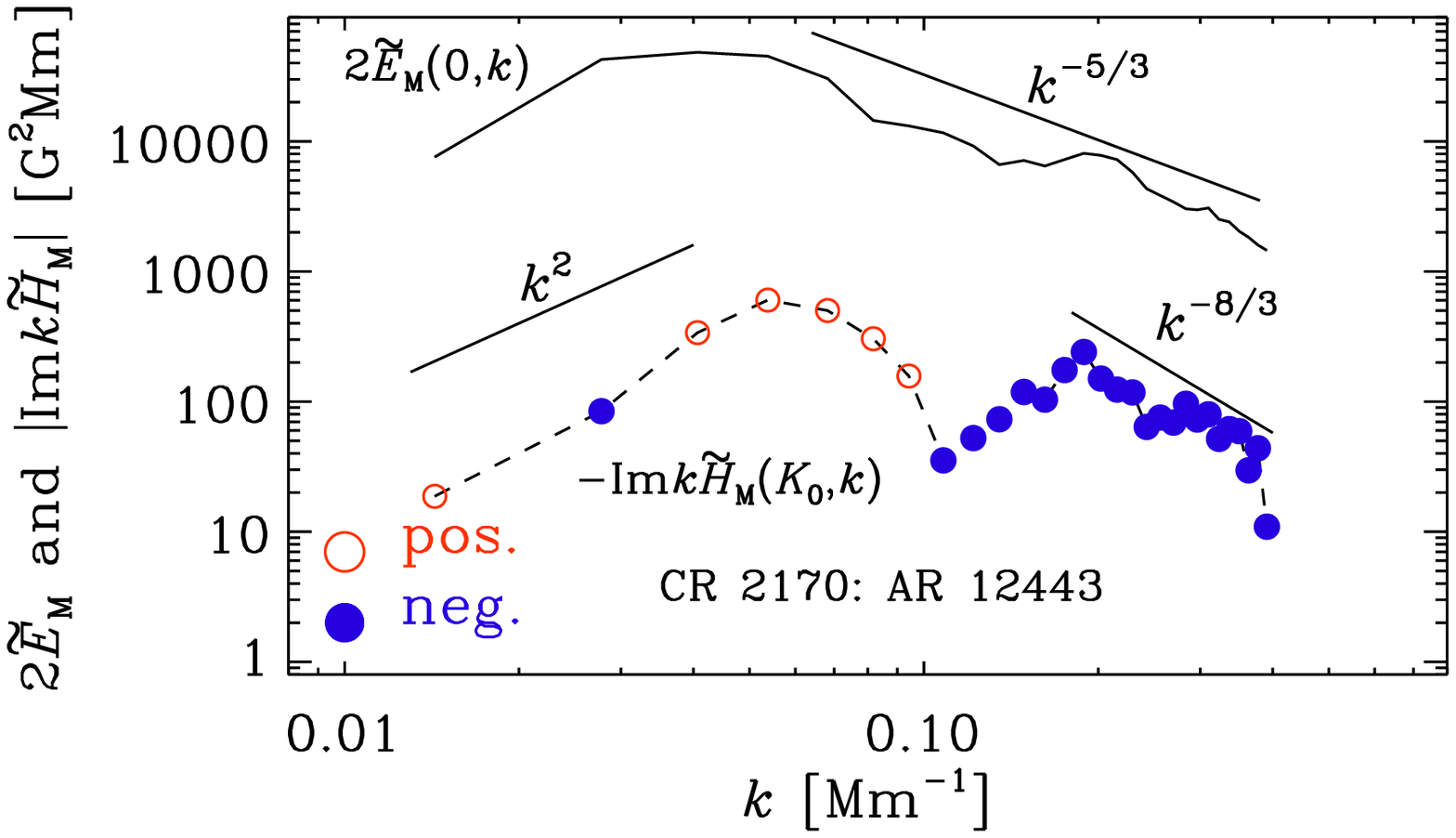}
\includegraphics[width=\columnwidth]{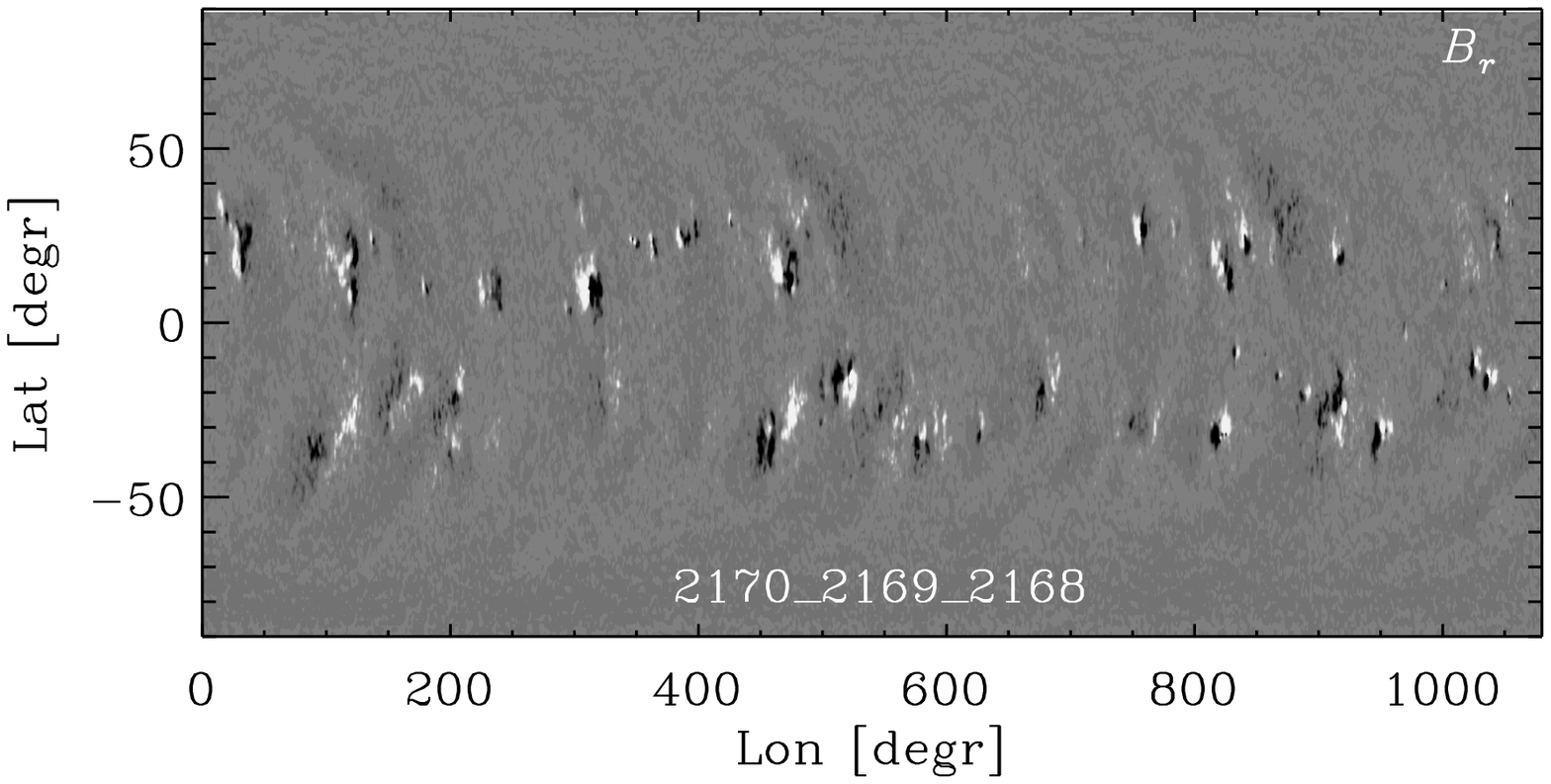}
\end{center}\caption[]{
Top and bottom panels show the same as in \Fig{2095-2093} for CRs 2168-2170.
The middle panel shows the spectra obtained from a smaller 2D patch
of size $80\degr \times 60\degr$ containing the AR~12443 that
emerged close to equator, $\vartheta=+6\degr$as a $\beta$-type
on 2015-10-30 during CR~2170.
}\label{2170-2168}
\end{figure}

\subsection{Late declining phase of SC~24}

During the later part of the declining phase,
the magnitudes of total magnetic energy and helicity
decrease, and the helicity sign shows fluctuations, as can be seen from
\Fig{tev_EHl}(b).
In \Fig{2170-2168}, we show that the spectrum of solar magnetic
helicity is very complex during this time epoch. It
shows multiple sign-reversals as a function of $k$.
During CRs 2168--2170,
the dominant sign of magnetic helicity in the north is positive,
thus violating the HSR. Moreover, sign changes at $k<0.03\Mm^{-1}$ reflect
possible fluctuations at the largest length scales, and it is not
necessarily caused by ARs.
These sign changes possibly occur due to spectral power being
proportional to $k^2$ expected for random fields that are
$\delta$-correlated in space.

However, the violation of the sign-rule seen at intermediate to large $k$
in the top panel of \Fig{2170-2168}
could indeed be caused due to emergence of some peculiar ARs.
To investigate this further, we focus on the AR~12443 that emerged
close to the equator during CR~2170. This developed a complex
$\beta\gamma\delta$ type structure, and gave rise to a couple of M-class
and several C-class flares. The spectra determined from a smaller
2D patch containing AR~12443 is shown in the middle panel
of \Fig{2170-2168}, demonstrating that this AR carries, unexpectedly, a net
positive magnetic helicity.
More dedicated numerical work is needed in this direction to explore
whether such sign anomalies are indeed associated with the morphological
complexities of ARs.
The proximity of this AR to the equator could be yet another possible reason
for the observed violation, as it is not clear if the underlying LSD
activity belt is strictly symmetric about the equator.

\begin{figure}\begin{center}
\includegraphics[width=\columnwidth]{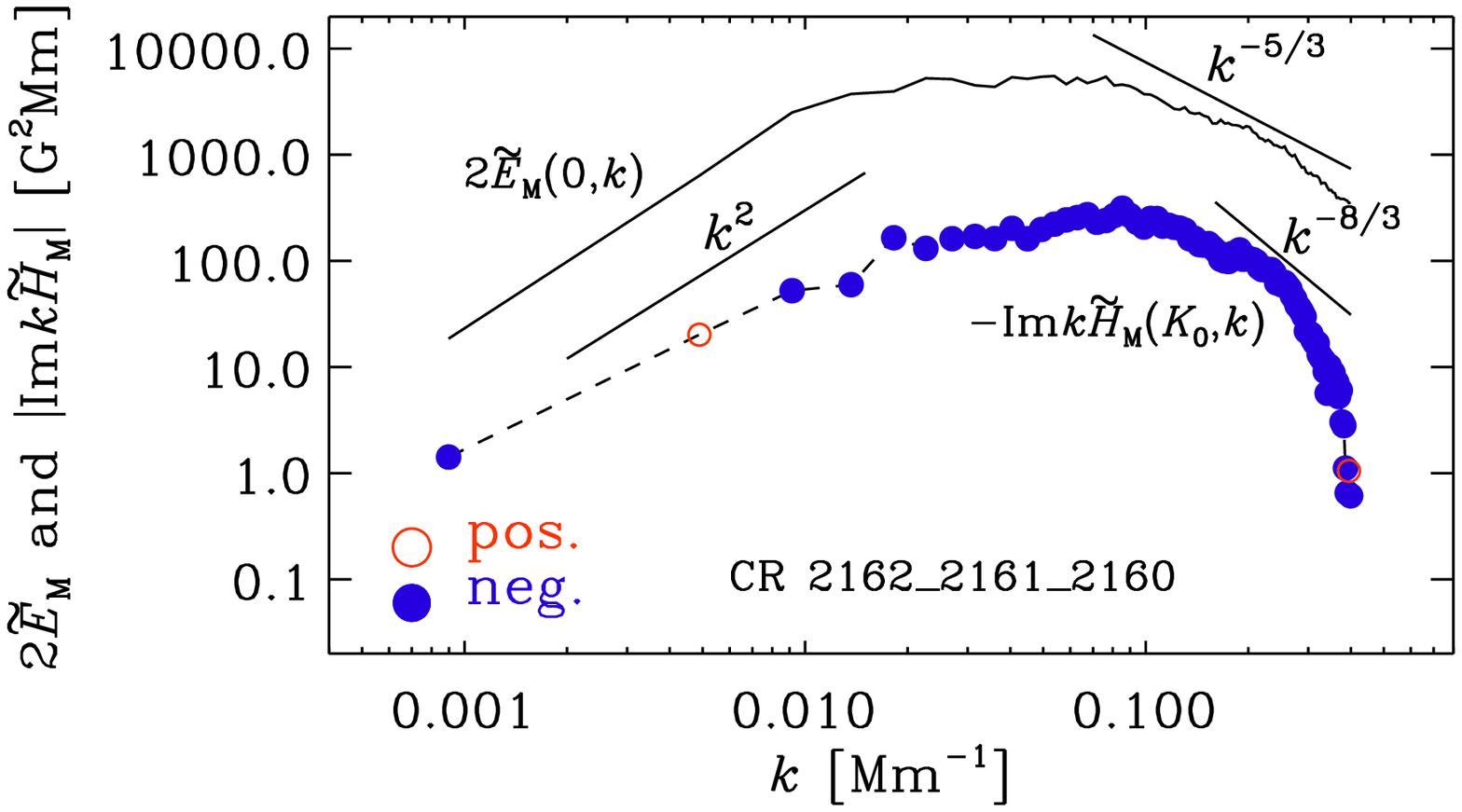}
\includegraphics[width=\columnwidth]{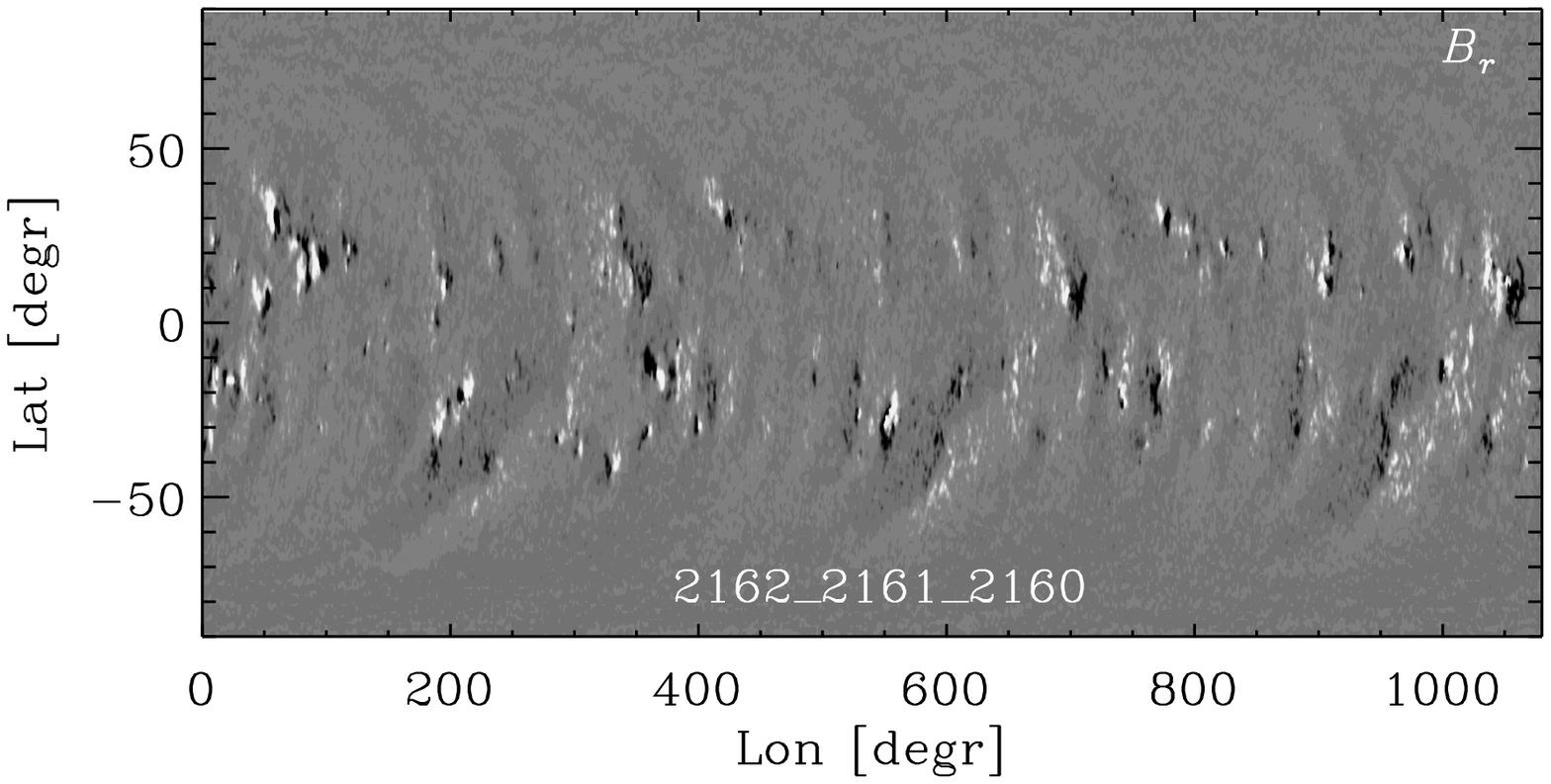}
\end{center}\caption[]{
Same as \Fig{2095-2093} but for the interval spanning CR~2160--2162.
BPS17 studied these same CRs using vector magnetograms data
obtained from SDO/HMI.
}\label{2162-2160}
\end{figure}

\subsection{Comparison to BPS17}
\label{comp}

In BPS17, {\em SDO}/HMI synoptic maps for CRs 2161--2163
were analyzed with an approach identical to that presented here.
The corresponding spectra from SOLIS data are shown in \Fig{2162-2160},
although the Carrington rotations included are not exactly matching those of the used
HMI data. The main difference is that no bihelical spectrum could be recovered
from the HMI data, while the SOLIS data shows a sign reversal. Also, the HMI data has
considerably higher spatial resolution, and therefore the data extends to far larger 
values of $k$ with better established powerlaws, while the SOLIS data fails to show
clear powerlaws. In both computations, we see fluctuations at the
largest wavenumbers possibly due to low amplitude small-scale magnetic
fields. 

\begin{figure}\begin{center}
\includegraphics[width=\columnwidth]{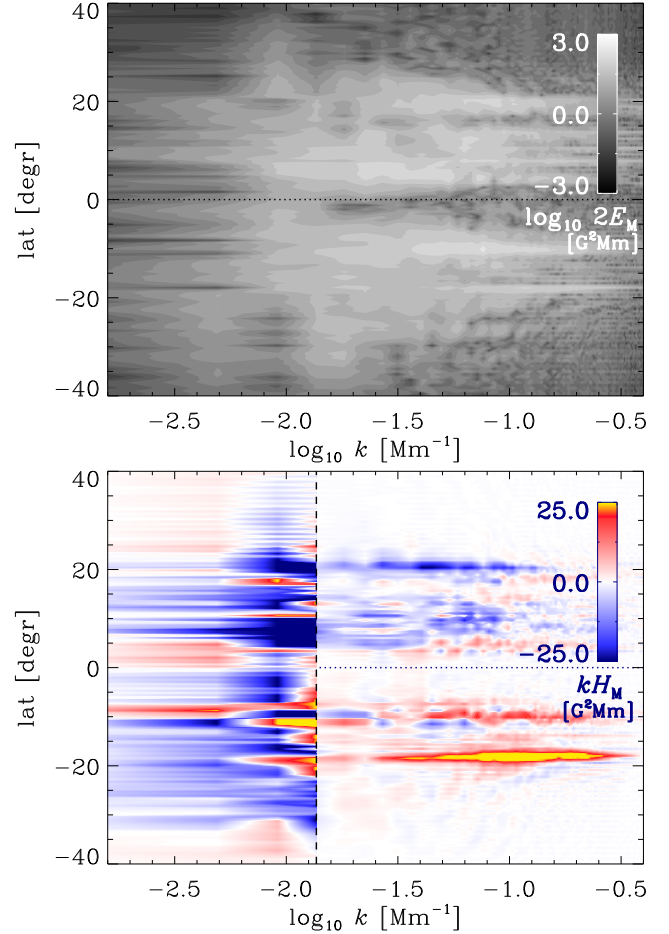}
\end{center}\caption[]{
Magnetic energy and helicity spectra for 2D solar surface data for
CR~2160--2162 as a function of $k$ and latitude.
Colors to the left of the
dashed vertical line are saturated at levels $\pm2\,{\rm G^2}\Mm$
to highlight the sign reversal in southern hemisphere.
}\label{pkHkNS_fullk_2162_2161_2160}
\end{figure}

To hunt down the reason for the difference seen at small wavenumbers,
we transform back from $K$ space to $Z$ space and show the spectra 
as a function of latitude in \Fig{pkHkNS_fullk_2162_2161_2160}. By
comparing this figure with Figure~9 of BPS17, we see that there is a
good agreement between the results at intermediate
scales. The retrieved extrema of the magnetic helicity are larger
for SOLIS than for HMI, with the magnetic energy values being in fair agreement.
At the small wavenumber end (largest scales), however, differences are obvious.
At the smallest $k$, the HMI data shows relatively strong signals extending to
high latitudes, violating the hemispheric sign rule especially in the high 
southern latitudes and northern lower latitudes. Similarly, the SOLIS signal
at the smallest $k$ comes from higher latitudes. The power at these scales
is lower than in the HMI spectra, and the sign rule is not violated as strongly.
In both the SOLIS and HMI data, the helicity and magnetic energy fade off
at high $k$, and both seem to indicate that north and south are somewhat asymmetric,
with the north decreasing more rapidly than the south. 
In conclusion, the intermediate scales seem to be in fair agreement
with both data sets, but some differences can be seen especially at the
largest scales.
Also, the sign change at low wavenumbers in the southern hemisphere
as apparent from \Fig{pkHkNS_fullk_2162_2161_2160} was not seen by BPS17.

Some of the discrepancies might relate to the differences between the two instruments.
SOLIS/VSM observations used for this analysis have a significantly higher signal-to-noise ratio
and spectral resolution compared to HMI \cite[see, e.g.,][]{JT12}.
Recent studies also show that SOLIS and HMI observations of meridional
and zonal magnetic fields often disagree at intermediate field strengths
(plage regions).
This most likely relates to fundamental limitations in Zeeman effect-based
vector magnetic field observations due to the unequal noise in the
transverse and line-of-sight components of the magnetic field.
HMI data processing also suffers from lack of a realistic filling factor, which most likely overestimates transverse fields outside active regions (private communication with SOLIS/VSM and SDO/HMI teams).

\section{Discussion}
\label{dc}

This study was motivated by the earlier paper by BPS17 where
no bihelical magnetic helicity spectra could be retrieved from HMI/SDO
data for three Carrington rotations during SC~24. Instead, the helicity spectrum
showed the same sign at large and small scales, inconsistent with
the expectations
from a helically driven $\alpha$ effect dynamo scenario. BPS17
argued that this might
simply be due to the epoch of the observations having been unfortunate. Another
line of thought was that the solar surface could be a special place
in between the
dynamo-active convection zone and the solar wind. These regions are
expected to show reversed signs
of helicities according to models \citep{WBM11,WBM12} and solar wind
observations \citep{BSBG11}, the sign change
possibly occurring in the surface regions, resulting in undetectable or weak
systematic helicity signatures.  Our current study shows that both lines
of thought were partially correct.
In fact, more realistic modeling now suggests that the sign change is expected to 
occur at a height of just $\sim5\Mm$ above the surface \citep{BSB18}.

Throughout the nearly seven years of data analyzed here, the power at
large scales is persistently
weaker than that in the mid-range scales, distinctively different from the dynamo simulations,
where the large-scales possess the largest power. This indicates that the helicity signatures
of the LSD are, indeed, weak near the surface, overwhelmed by the
helicity signal that the active regions carry, influenced by the SSD, and
perhaps most importantly, prone to be affected by noise and any uncertainties related to the
data analysis procedures.
Our analysis reveals that the expected bihelical signature
can be retrieved easily from time-averaged spectra
as computed from the high signal-to-noise SOLIS/VSM synoptic maps,
but it also highlights the need for better synoptic maps,
covering a significant fraction of the SC, allowing us
to the finding of opposite sign of helicity at large scale as compared
to results in BPS17; see \Sec{comp}.

We recover a rather weak dependence on the SC, but certain patterns can be
discerned. The probability of recovering a bihelical, hemispheric sign rule-obeying spectrum
is increased during the rising phase of the SC. Magnetic helicity tends to maximize
not during the sunspot maximum, but after some delay, and the descending phase is
characterized with almost random kinds of helicity spectra. 
During the solar minimum we
observe an increased probability to find 
HSR violating helicity spectra. These findings
are in partial agreement with earlier work (sign change in between
the ascending and descending
phases), which has been reported before 
\citep[e.g.][]{BBS03}  and for which also theoretical explanations
have been proposed \citep[e.g.][]{CCN04,Pipin2013}. 
Inexplicable features in our data (the reversed sign
also at the large scales, the highly variable behavior during the descending phase), however, remain.

One scenario that could explain 
the highly variable behavior in the descending phase are
contributions arising from very complex active regions. In this work, we analyzed
only one such region, but we were able to show that such a region can contribute
significantly to reversed helicity sign at intermediate scales. Their relative abundance
to less complex active regions is known to be elevated during the descending phase
of the SC \citep{JN16}.
Another possibility could be that signals from ARs occurring close
to equator might leak into the opposite hemisphere, thus polluting
the spectrum from this hemisphere with the wrong sign.
This scenario does not, however, explain the reversed signs
of helicity at large scales.

As discussed in \Sec{edp}, the magnetic helicity spectra 
obtained during the early declining phase
show intriguing features of clear HSR-obeying bihelicity, large power at the
small wavenumbers, together with a Kazantsev spectral slope at large scales.
It is not obvious if systems with magnetic Prandtl number
$\Pem=\nu/\eta\ll1$ with $\nu$ and $\eta$ being the kinematic
viscosity and microscopic resistivity, respectively, must always
host an SSD which is much harder to excite in such a regime, making
it somewhat an open issue whether the Sun, being a low-$\Pem$ object,
indeed supports SSD.
However, although dynamos at $\Pem<1$ are harder to excite \citep{Scheko05},
the adverse excitation conditions at $\Pem=0.1$ are now understood to be
a consequence of the bottleneck effect in turbulence \citep{Isk07}.
This effect is particularly strong when turbulence is forced at the
scale of the domain.
Simulations of \cite{SB14} at larger forcing wavenumbers resulted in no
visible increase of the critical dynamo number.
Once the dynamo is excited, the bottleneck effect is suppressed, so
the low-$\Pem$ controversy is hardly relevant in the nonlinear regime
\citep{Bra14}.

Based on the Kazantsev spectrum seen in \Fig{max2156_spec} and
bearing in mind the discussion of the previous paragraph, we note that
these results are suggestive of both LSD and SSD being operative
simultaneously in the Sun.
It remains to be seen how it all fits into a unified scheme of
SSDs and LSDs such as the one explored by \cite{Sub99}.
More numerical works covering a sufficiently broad range of scales
is needed in this interesting but complicated regime.

\section{Conclusions}
\label{co}

The present work has shown that the solar magnetic fields are bihelical,
best observed during maximum activity of the Sun,
with opposite signs of magnetic helicity at large and small
length scales---exactly as expected from a helically driven global
solar dynamo. Nearly $75\%$ of all the analyzed synoptic maps show
agreement with the HSR in terms of net magnetic helicity, which is
dominated by ARs and thus becoming negative (positive) in the northern
(southern) hemisphere. In agreement with some previous claims, the violations
of the HSR is mostly seen during the early rising phase of the SC.

We have also highlighted the need for more reliable and better data, as
it is possible that it is not the Sun, but the data itself that is
more enigmatic, leading to opposite claims based on measurements
from different instruments. We discussed one such example while noting
some more from the literature.
Therefore, improved data quality from upcoming missions such as
\emph{Solar Orbiter} with synergetic measurements from other facilities
like DKIST, is critical to establishing some fundamental claims
about the solar helicity.

\begin{acknowledgements}
This work utilizes SOLIS data obtained by the NSO Integrated Synoptic
Program (NISP), managed by the National Solar Observatory, which is
operated by the Association of Universities for Research in Astronomy
(AURA), Inc.\ under a cooperative agreement with the National Science
Foundation.
Financial support from the Academy of Finland grant No. 272157 to
the ReSoLVE Centre of Excellence (MJK, PJK, and IV) and through the
Max-Planck-Princeton Center for Plasmaphysics (NS) is gratefully acknowledged.
This research was further supported by the NSF Astronomy and Astrophysics
Grants Program (grant 1615100), and the University of Colorado through
its support of the George Ellery Hale visiting faculty appointment (AB).
\end{acknowledgements}


\end{document}